\documentclass[natbib]{svjour3} 
\smartqed 
\usepackage{graphicx}
\usepackage{mathptmx}
\usepackage{natbib}
\usepackage{multirow}
\usepackage{svn}

\usepackage{bm}
\usepackage{amssymb}

\newcommand{\cC}{\mathcal{C}}
\newcommand{\cH}{\mathcal{H}}
\newcommand{\cG}{\mathcal{G}}
\newcommand{\cl}{\mathcal{L}}
\newcommand{\bell}{\hat{\hspace{-.05cm}{\boldsymbol \ell}}}
\newcommand{\bxi}{{\bm \xi}}
\newcommand{\bzeta}{{\bm \zeta}}
\newcommand{\bbe}{{\mathbf e}}
\newcommand{\bg}{{\mathbf g}}
\newcommand{\br}{{\mathbf r}}
\newcommand{\brs}{{\mathbf r}_{\rm s}}
\newcommand{\bbs}{{\mathbf s}}

\newcommand{\bv}{{\mathbf v}}
\newcommand{\nab}{{\bm \nabla}}

\newcommand{\xb}{{\mathbf x}}
\newcommand{\bz}{{\mathbf z}}
\newcommand{\bk}{{\mathbf k}}
\newcommand{\bG}{{\mathbf G}}
\newcommand{\bS}{{\mathbf S}}
\newcommand{\bkhat}{\hat {\mathbf k}}
\newcommand{\de}{\delta}

\newcommand{\om}{\omega}

\newcommand{\Ga}{\Gamma}

\newcommand{\ts}{t_{\rm s}}
\newcommand{\zs}{z_{\rm s}}

\newcommand{\x}{\nonumber \\}
\newcommand{\bX}{{\mathbf X}}

\newcommand{\thedelta}{16.0347} 
\newcommand{\dtmn}{\delta\tau_{\rm mn}}
\newcommand{\aap}{{Astron. Astrophys.}}
\newcommand{\apj}{{Astrophys. J.}}
\newcommand{\apjl}{{Astrophys. J.}}

\newcommand{\solphys}{{Solar Phys.}}
\newcommand{\nat}{{Nature}}
\newcommand{\araa}{{Annu. Rev. Astron. Astrophys.}}
\newcommand{\mnras}{{Month. Not. Roy. Astron. Soc.}}

\begin{document}

\title{Interpretation of Helioseismic Traveltimes
}
\subtitle{Sensitivity to sound speed, pressure, density, and flows}


\author{Raymond Burston \and Laurent Gizon \and Aaron~C.~Birch}

\institute{
R. Burston, L. Gizon, A.~C.~Birch\at
              Max-Planck-Institut f\"ur Sonnensystemforschung, Justus-von-Liebig-Weg 3, 37077 G\"ottingen, Germany\\
              \email{gizon@mps.mpg.de}
           \and
            L. Gizon \at
              Institut f\"ur Astrophysik, Georg-August-Universit\"at G\"ottingen, Friedrich-Hund-Platz 1, 37077 G\"ottingen, Germany\\
}

\date{Received: date / Accepted: date}

\maketitle

\begin{abstract}
Time-distance helioseismology uses cross-covariances of wave motions on the solar surface to determine the travel times of wave packets moving from one surface location to another. We review the methodology to interpret travel-time measurements in terms of small, localized perturbations to a horizontally homogeneous reference solar model.  Using the first Born approximation, we derive and compute 3D travel-time sensitivity (Fr\'echet)  kernels for perturbations in sound-speed, density, pressure, and vector flows. While kernels for sound speed and flows had been computed previously, here we extend the calculation to kernels for density and pressure, hence providing a complete description of the effects of solar dynamics and structure on travel times. We treat three thermodynamic quantities as independent and do not assume hydrostatic equilibrium. We present a convenient approach to computing damped Green's functions using a normal-mode summation. The Green's function must be computed on a wavenumber grid that has sufficient resolution to resolve the longest lived modes. The typical kernel calculations used in this paper are computer intensive and require on the order of 600 CPU hours per kernel. Kernels are validated by computing the travel-time perturbation that results from horizontally-invariant perturbations using two independent approaches. At fixed sound-speed, the density and pressure kernels are approximately related through a negative multiplicative factor, therefore implying that perturbations in density and pressure are difficult to disentangle. Mean travel-times are not only sensitive to sound-speed, density and pressure perturbations, but also to flows, especially vertical flows. Accurate sensitivity kernels are needed to interpret complex flow patterns such as convection.
\keywords{Helioseismology \and Solar interior  \and Wave propagation \and Scattering }
\end{abstract}

\newpage

\tableofcontents

\section{The forward problem in time-distance helioseismology}

There are two general approaches to helioseismology. Global helioseismology consists of interpreting the frequencies of the normal modes of oscillations, whereas local helioseismology uses spatial-temporal correlations of the wave field as the input data (time-distance helioseismology, acoustic holography, direct modeling) or local power spectra of solar oscillations (ring-diagram analysis). For overviews of local helioseismology techniques, see reviews by \cite{2002RvMP...74.1073C}, \cite{2005LRSP....2....6G}, and \cite{2010ARA&A..48..289G}.

In this paper, we focus on the forward problem of time-distance helioseismology, i.e. how to compute the effects of subsurface heterogeneities on wave travel times. The problem of inferring subsurface heterogeneities from observed travel times (the inverse problem) was first formalised and solved by \cite{1996ApJ...461L..55K} and is also discussed in the reviews cited above. Throughout this paper we consider travel times measured between two points on the solar surface using a cross-covariance technique \citep{1993Natur.362..430D}. The purpose of the paper is to review and extend the computation of the sensitivity of wave travel times to small-amplitude changes to the local physical conditions of the solar interior. By a small-amplitude change to a reference solar model we mean a perturbation that causes a phase shift in the observed wave field that is much less than a radian. These sensitivity functions depend on the 3D position in the solar interior (the coordinates of local scatterers) and are known as Fr\'echet kernels, or simply kernels, in Earth seismology  \citep{1998GeoJI.132..521M,1999GeoJI.137..805M,2000GeoJI.141..157D,2000GeoJI.141..175H}. The solar problem differs from the earthquake seismology problem because solar oscillations are continuously excited by a random process, turbulent solar convection (granulation).

The first kernel calculations in time-distance helioseismology employed the ray approximation, in which the wavelength is assumed to be small compared to the other length scales in the problem \citep{1996ApJ...461L..55K,1997ASSL..225..241K}. \cite{2000SoPh..192..193B} used the first Born approximation (single-scattering approximation) to compute 3D kernels for sound-speed perturbations using the single-source approximation. Subsequently, \cite{2002ApJ...571..966G} introduced a general framework for Born kernel calculations including a treatment of wave excitation by near-surface convection, this framework is the basis for the calculations presented here. In particular, \cite{2002ApJ...571..966G} showed that spatially distributed wave sources alter the sensitivity kernels: scattering far away from the observation points can be important. They also showed that the details of the wave power spectrum, wave attenuation, and the details of the measurement procedure must all be included in the kernel computations. They proposed a measurement procedure that enables a consistent computation of kernels by relating small-amplitude changes in the cross-covariance to travel-time shifts. Using this approach kernels have been computed for source amplitude and wave attenuation \citep{2002ApJ...571..966G}, sound-speed perturbations \citep{2004ApJ...608..580B}, and vector flows \citep{2000JApA...21..339G,2006ESASP.617E..38J, 2007ApJ...671.1051J,2007AN....328..228B, 2011AN....332..658B}. We note that the same approach was used to compute kernels for ring-diagram analysis \citep{2007ApJ...662..730B}. All of these calculations assume that the helioseismic observable is the vertical wave velocity at the solar surface. Away from disk center, the horizontal velocity contributes to the Doppler velocity and short wavelengths are less visible towards the limb \citep{2006ESASP.617E..38J, 2007ApJ...671.1051J}. All of these calculations employed a plane-parallel approximation for the background solar model (used for the study of small patches on the Sun).

More recent approaches based on the adjoint method are particularly useful for solving problems with 3D heterogeneous background models \citep{2011ApJ...738..100H}. This method has been applied to a background model containing a magnetic field and 3D kernels for sound-speed, density, 3D flows, and magnetic fields were computed \citep{2012PhRvL.109j1101H}. Mathematical expressions for density kernels, as well as sound-speed kernels, have also been presented by \cite{2011arXiv1105.4619S} using ``functional analytic methods". The contribution of the lowest-order non-zero perturbations of magnetic field, to a background containing no magnetic field, is discussed in \citet{2007AN....328..204G}.

Other approaches than the Born approximation exist.  For example, \citet{2003A&A...412..257J} applied the Rytov approximation, which is closely related to the Born approximation, to compute kernels for sound-speed and flows in the single-source approximation.

In this paper, we work within the framework of \cite{2002ApJ...571..966G} and extend the computation of travel-time kernels to treat density and pressure perturbations, in addition to sound-speed and flow  perturbations. For the sake of computational simplicity, we work in Cartesian geometry starting from a 1D static and non-magnetic reference solar model. Furthermore, we do not assume {\it a priori} hydrostatic equilibrium in the perturbed model to allow for contributions to the force balance from flows and magnetic fields and thus we retain three thermodynamic variables (sound speed, density, and pressure in the perturbed model).

\section{Equivalent-source description of wave interaction}\label{seq.equiv_source_desc}

Linear adiabatic stellar oscillations \citep[e.g.][]{1967MNRAS.136..293L} are typically governed by equations of the form $\cl \, \bxi(\br,t) = \bS(\br,t)$, where $\cl$ is a linear differential operator, $\bxi(\br,t)$ is the vector-valued wave displacement, and $\bS(\br,t)$ is the vector-valued source function responsible for wave excitation. Scattering events are uniquely identified by both time $t$ and a three dimensional (3D) position vector $\br$.

In the first-order Born approximation \citep[e.g.][]{2002ApJ...571..966G}, the standard coupled system of linear differential equations is
\begin{eqnarray}
\cl_0 \, \bxi_0 &=& \bS_0,\label{Lxi_eq_S} \\
\cl_0 \,\delta \bxi &=&-\delta\cl \,\bxi_0\, + \delta \bS. \label{eneqw}
\end{eqnarray}
Equation (\ref{Lxi_eq_S}) describes the propagation of small amplitude waves through a background solar model. The operator $\cl_0$ is the reference linear wave operator, $\bxi_0$ the displacement field in the reference model and $\bS_0$ is the reference source field. Equation (\ref{eneqw}) governs the perturbed displacement field, $\delta \bxi$. The first-order linear operator $\de \cl$ encapsulates perturbations to the solar thermodynamic structure and the introduction of flows, and $\de \bS$ is the perturbation to the source field. In this paper, we ignore $\delta \bS$ \citep[see e.g.][for a discussion]{2002ApJ...571..966G}.

For a reference solar model with vanishing flows and magnetic fields, and also using the Cowling approximation, the reference linear operator governing the reference displacement field is \citep[see e.g.][]{2006ESASP.624E..63C}
\begin{eqnarray}\label{L0}
\cl_0\, \bxi_0 &:=& \rho_0 \, \left(\partial_t+\Gamma\right)^2\bxi_0- \nab \left(c_0^2\, \rho_0\, \nab\cdot \bxi_0 + \bxi_0\cdot\nab P_0\right) +\bg_0\, \nab \cdot \left(\rho_0 \,\bxi_0 \right).
\end{eqnarray}
The background solar model variables, which depend on the distance from the center of the Sun, have their usual symbols: $c_0$ is the sound-speed, $\rho_0$ is the density, $P_0$ is the pressure, and $\bg_0$ is the acceleration due to gravity. The partial derivative with subscript $t$ denotes differentiation with respect to time and the vector $\nab$ is the 3 dimensional spatial gradient operator. The phenomenological operator, $\Gamma$, is responsible for damping the oscillations. The random source function $\bS_0$ is assumed to be statistically homogeneous in the horizontal direction and statistically stationary in time and represents the driving of oscillations by near-surface turbulent convection.

The first-order linear operator $\de \cl$ due to first-order perturbations in sound-speed squared ($\delta c^2$), density ($\delta \rho$), pressure ($\delta P$) and 3D mass flows ($\bv$) is,
\begin{eqnarray}\label{deltaLxi}
\de\cl \bxi_0 :=&&- \nab\left(\de c^2\, \rho_0\, \nab\cdot \bxi_0\right) + 2 \, \rho_0\,(\bv\cdot\nab)\partial_t \, \bxi_0- \nab\left(\bxi_0\cdot\nab \de P\right) \x
&&+ \de \rho \,\left(\partial_t+\Gamma\right)^2\bxi_0 - \nab\left(c_0^2\, \de\rho\, \nab\cdot \bxi_0\right)+\bg_0\, \nab \cdot(\de\rho \,\bxi_0) +\delta\bg\,\, \nab \cdot(\rho_0 \,\bxi_0) .
\end{eqnarray}
The first-order changes in gravity $\delta\bg$ are related to first-order changes in density and are expected to be very small near the solar surface.

We use sound-speed, density, and pressure as our three independent thermodynamic variables. It is possible to make other choices; e.g. the adiabatic exponent $\gamma$ may be introduced at the expense of the others by using the relationship $\rho \, c^2 = P \, \gamma$. We do not assume hydrostatic equilibrium for the thermodynamic perturbations. If hydrostatic equilibrium was assumed, then it would be possible to eliminate the gradient of the pressure perturbation (in Eq.~[\ref{deltaLxi}]) with a contribution from the density perturbation (or vice-versa), and consequently, reduce the number of independent perturbation variables to two (e.g. sound-speed and density) as is done in global helioseismology \citep[e.g.][]{2002RvMP...74.1073C}.

\section{Computing travel-time sensitivity kernels for structure and flows}

In time-distance helioseismology wave travel times ($\tau$) are measured from the temporal cross-covariance of the observed signal, $\phi(\xb,t)$:
\begin{equation}\label{eq.xcorr_def}
C(t; \xb_1,\xb_2) = \frac 1T \int_0^T dt^\prime \phi(\xb_1,t^\prime) \phi(\xb_2,t^\prime+t),
\end{equation}
where $t$ is the time lag, $\xb_1$ and $\xb_2$ are two spatial locations on the solar surface and $T$ is the duration of the time series. In the temporal Fourier domain it becomes
\begin{equation}\label{eq.xcorr_def_fourier}
C(\omega; \xb_1,\xb_2) = h_\omega \,\, \phi^*(\xb_1,\omega) \phi(\xb_2,\omega),
\end{equation}
where $\omega$ is the angular frequency and the superscript $*$ denotes complex conjugate \citep[throughout this paper we use the Fourier conventions from][]{2002ApJ...571..966G}, $C(\omega; \xb_1,\xb_2)$ is the temporal Fourier transform of $C(t; \xb_1,\xb_2)$ and $\phi(\xb,\omega)$ is the temporal Fourier transform of $\phi(\xb,t)$. The sampling in frequency space is inversely proportional to the duration $h_\omega= 2 \pi / T$.

The observable $\phi$ is often a line-of-sight Doppler velocity; it is connected to the oscillations of the atmosphere caused by  seismic waves. There exists a non-trivial relationship between $\bxi$ and $\phi$ \citep[see e.g.][]{2014arXiv1401.3182N} which can be expressed in general as
\begin{equation}
\phi(\xb,\omega)= O \left\{ \bxi(\br,\omega) \right\},
\end{equation}
where $O$ is an operator that acts on the 3D displacement and maps it onto the CCD detector. The operator includes the Point Spread Function (PSF) of the instrument plus any filters that were applied to the data.

Then using the Born approximation (Section \ref{seq.equiv_source_desc}) the first-order change to the cross-covariance is given by
\begin{eqnarray}
&&\delta C (\omega; \xb_2,\xb_1)   = -(2\pi)^4 \sum_\alpha \int_{\odot} d\br \sum_j \cG^j(\xb_2,\omega, \br)    \;\; \hat{\mathbf{e}}_j \cdot \delta\mathcal{L}^{\alpha}_{\br,\omega}  \,\,\bX (\br, \omega, \xb_1) + (1\leftrightarrow 2)^*, \,\,\,\,\,\,\,\,\,\,\,\,\,\,     \label{eq.delta_xcorr1} \\
&&\qquad\qquad\,\,\,\,\,\,\,\,\,\,\,\,\,\,\,  = \sum_\alpha \int_{\odot} d\br \,\, \cC_\alpha (\br, \omega; \xb_2,\xb_1) \delta q_\alpha(\br)+ (1\leftrightarrow 2)^*,\label{eq.delta_xcorr2} \\
&&{\rm where} \quad \bX(\br, \omega, \xb_1)         :=   \sum_{i,i'}\int_{\odot} d\br_s  \int_{\odot} d\br_s' \cG^{i*}(\xb_1,\omega, \br_s)  \bG^{i'}(\br,\omega,\br_s')   M_{ii'}(\br_s,\br_s', \omega), \label{eq.X_def} \\
&&{\rm and}\quad \,\,\,\,\,\,M_{ii'}(\br_s,\br_s', \omega)  :=    E[ S_i^*(\br_s,\omega)  S_{i'}(\br_s',\omega)] .
\end{eqnarray}
The Green's tensor $ \bG^i$ is defined by
\begin{eqnarray}
\cl_0\,\bG^i(\br,t - \ts,\br_s) = \de(\br -\brs) \,\de(t-\ts) \, \hat{\bbe}_i  ,
\end{eqnarray}
and for convenience we also defined
\begin{eqnarray}
\cG^i &=&O \left\{  \bG^i\right\}.
\end{eqnarray}
The operator $\delta\mathcal{L}^{\alpha}_{\br,\omega}$ is the component of the perturbed wave operator ($\delta\mathcal{L}$) that is due to the change in the physical quantity $\delta q_\alpha(\br)$. The quantity $M_{ii'}$ is the source covariance tensor. The symbol $(1\leftrightarrow 2)^*$ is a short notation to mean switch $\xb_1$ and $\xb_2$ and take the complex conjugate of all other terms in the expression. The scattering physics is included in the functions $\cC_\alpha$ that give the linear sensitivity of the expectation value of the cross-covariance to changes in the solar model. In order to arrive at Equation (\ref{eq.delta_xcorr2}), it is often required to use integration-by-parts to separate the perturbation variables from spatial derivatives, and thus in those cases it is been assumed that relevant quantities vanish on the boundaries.

Using Equation (\ref{eq.delta_xcorr2}),  $\cC_\alpha$ can be expressed for each of the perturbations
\begin{eqnarray}
\cC_{c^2} (\br; \xb_2,\xb_1)  &=&  -(2\pi)^4\rho_0(\br) \, c_0^2(\br) \, \left[\sum_j  \partial_{r_j} \cG^j(\xb_2, \br)  \right]\, \nab_{\br} \cdot \bX (\br, \xb_1)  + (1\leftrightarrow 2)^*   ,\,\,\,\,\,\,\,\,   \\
\cC_{\rho} (\br; \xb_2,\xb_1)  &=&-(2\pi)^4 \rho_0(\br) \, \Big[ (\omega +{\rm i} \gamma)^2  \, \sum_j   \cG^j(\xb_2, \br)  X_j (\br, \xb_1)  \x
&&+ c_0^2(\br)\, \nab_{\br} \cdot \bX (\br, \xb_1)  \,\sum_j  \partial_{r_j}  \cG^j(\xb_2, \br)     \x
&&-  \bX (\br, \xb_1)  \cdot \nab_{\br} \,\sum_j   \cG^j(\xb_2, \br) \, g_j(\br)  \Big]\x
&&-(2\pi)^4 \delta g_i(\br) \, \frac{\rho_0}{\delta \rho}(\br)\,\cG^i(\xb_2, \br)  \, \nab_{\br} \cdot \left[\rho_0 \bX (\br, \xb_1) \right] + (1\leftrightarrow 2)^* , \label{eq.def_curlyc_rho}\\
\cC_{P} (\br;\xb_2,\xb_1)     &=&  (2\pi)^4 P_0(\br) \,\nab \cdot  \left[ \bX (\br, \xb_1) \sum_j  \partial_{r_j}  \cG^j(\xb_2, \br)  \right]  + (1\leftrightarrow 2)^*   ,\\
\cC_{\bv_i} (\br; \xb_2,\xb_1)     &=&  (2\pi)^4{\rm i} \,2 \,\omega\, \rho_0(\br) \,\cG^j(\xb_2, \br)  \partial_{r_i} X_j (\br, \xb_1) + (1\leftrightarrow 2)^* ,
\end{eqnarray}
where $\rm i$ is the imaginary number. Note that the first-order perturbation to gravity can be expressed in terms of first-order density according to
\begin{eqnarray}
\nab \cdot \delta \bg = -4 \pi \, G \, \delta \rho
\end{eqnarray}
where $G$ is the gravitational constant.

The integrals in Equations (\ref{eq.delta_xcorr1})-(\ref{eq.X_def}) are 3D and over the volume of the Sun. The 3D flow vector ($\bv$) has been decomposed into its three components denoted $v_i$. The sensitivity kernel $K_\alpha$ for a given physical quantity $q_\alpha$  is computed with all other physical quantities held fixed. Each kernel $K_\alpha(\br)$ due to a point scatterer at $\br$ in the interior is computed according to
 \begin{eqnarray}
K_\alpha(\br;\xb_1,\xb_2) &=& 4\pi \, {\rm Re}\int_0^\infty \hspace{-.15cm} W^*(\om;\xb_1,\xb_2)\, \cC_\alpha (\br, \om; \xb_1, \xb_2) \, d\om,
\end{eqnarray}
where ${\rm Re}$ takes the real part of the following expression and the function $W$ gives the linear dependence of the travel-time shifts on the cross-covariance. The functions $W$ incorporate all the details of the travel-time measurement procedure \citep{2002ApJ...571..966G}. One-way travel-times are sensitive to all types of perturbations. \cite{1997SoPh..170...63D} introduced travel-time differences $\delta \tau_{\rm diff} = \delta\tau(\xb_1,\xb_2)- \delta\tau(\xb_2,\xb_1)$ to partially isolate the contribution from flows and mean travel times $\delta \tau_{\rm mean} = [ \delta\tau(\xb_1,\xb_2)+ \delta\tau(\xb_2,\xb_1)]/2$ to measure the other contributions. The corresponding $W$-functions are
\begin{eqnarray}
W_{\rm diff}(\omega; \xb_1,\xb_2) &=& W(\omega; \xb_1,\xb_2) - W(\omega; \xb_2,\xb_1)  ,\\
W_{\rm mean}(\omega; \xb_1,\xb_2) &=& [ W(\omega; \xb_1,\xb_2) + W(\omega; \xb_2,\xb_1) ]/2 .
\end{eqnarray}

The forward problem consists of modeling the chain of perturbations $\delta \bxi \rightarrow \delta\phi \rightarrow \delta C \rightarrow \delta \tau$ caused by changes in the structure and dynamics of the solar interior. Travel-time shifts ($\delta\tau)$ due to first-order Born perturbations in sound-speed squared, density, pressure and 3D vector flows are given by
\begin{eqnarray}
\delta\tau (\xb_1,\xb_2) &=& \int_\odot  d\br \left[ K_{c^2}(\br;\xb_1,\xb_2) \, \frac{\delta c^2}{c^2_0}(\br)  +  K_{\rho}(\br;\xb_1,\xb_2) \, \frac{\delta \rho}{\rho_0}(\br)  \right. \x
&&\left. \qquad \,\,\,\,\,  +   K_{P}(\br;\xb_1,\xb_2) \, \frac{\delta P}{P_0}(\br)  +  \sum_{i=1}^3 K_{v_i}(\br;\xb_1,\xb_2) \,  v_i(\br)  \right] .
\label{eq.dt_from_dq}
\end{eqnarray}

\section{Plane parallel example}

Here we compute example kernels assuming a plane-parallel model which is appropriate for studying small areas on the Sun. We assume the observable is given by
\begin{equation}
\phi(\bk,\omega) = -{\rm i} \omega \, F(\bk,\omega) \, \bell \cdot \bxi(\bk,z_{\rm obs},\omega),
\end{equation}
where $\bk$ is a horizontal wave vector, $\bell$ is the line-of-sight unit vector that we assume points in the vertical direction ($\bell = \hat{\mathbf z}$). We used an observation height of $z_{\rm obs}=300$ km. 

The function $F$ is a combination of both the instrumental PSF and filters. In this paper, we focus on modeling observations of waves with the same radial order. The solar power spectrum shows ridges of enhanced power. Each of these ridges corresponds to waves with a particular radial order. To isolate a single radial order, the power spectrum is filtered using a ridge filter that suppress all waves at other radial orders \citep[e.g. see][]{2008JPhCS.118a2033J}. In this paper we consider both an f-mode ridge filter and $p_1$-mode ridge filter.

We assume spatially uncorrelated sources, such that the example source covariance tensor is
\begin{equation}
M_{i i'}(\br_s,\br_s', \omega) = (2\pi)^{2} \delta_{i, z} \delta_{i', z}\delta(\xb_s-\xb'_s) m(\omega) \partial_{z_s} \delta(z_s-z_{\rm src})   \partial_{z_s'} \delta(z_s'-z_{\rm src}) ,
\end{equation}
where $\br_s=(\xb_s, z_s)$ and $\br_s'=(\xb_s', z_s')$.
This is the source covariance described by \cite{2004ApJ...608..580B}. The depth dependence is a vertical derivative of a vertical momentum source. This form of the source covariance gives a good match to the observed power spectrum. The frequency dependence here is Gaussian as in \cite{2004ApJ...608..580B} however, there is an additional factor of $1/\omega^2$ since we are solving for displacement whereas \cite{2004ApJ...608..580B} was solving for velocity. Here, the source auto-correlation function is $m(\omega) := \omega^{-2}  \exp[ - (\omega T_{{\rm src}})^2]$,  with a correlation time $T_{{\rm src}}=48$~s. We note that $M$ is defined to within an overall scale  factor, which changes the amplitude of the $\cC_q$ but does not change the kernels $K_q$ since the $W$-functions are inversely proportional to this scale factor. We used a source height of $z_{\rm src}=-100$ km \citep{2004ApJ...608..580B}.

\subsection{Convenient mode-summation formula for the Green's tensor}
\label{sec.greens}

The solutions to Equations (\ref{Lxi_eq_S}) and (\ref{eneqw}) can be written in terms of Green's functions \citep{2002ApJ...571..966G}. We use normal-mode summation to compute Green's functions, however, the approach presented here is different to what was implemented in \cite{2004ApJ...608..580B}. Here we write the Green's functions for the damped problem in terms of normal modes of the undamped problem, and consequently, the new approach is less mathematically and computationally demanding. The equation governing the reference solar oscillations can be expressed as
\begin{eqnarray}
\cl \bxi_0 = \rho_0 \left[ (\partial_t +\Ga)^2 +\cH \right]\bxi_0 = \bS_0,
\end{eqnarray}
where $\cH$ can be identified from Section \ref{seq.equiv_source_desc}. Normal-mode eigensolutions (undamped) thus must satisfy
\begin{eqnarray}
\cH \bxi^n(\br, \bk) = \om_n^2(k) \, \bxi^n(\br, \bk),
\end{eqnarray}
where $\bxi^n(\br, \bk)$ is the eigenfunction with radial order $n$ and horizontal wave vector $\bk$, $\omega_n(k)$ is the associated (real) eigenfrequency, and $k := || \bk ||$ is the wavenumber. Notice that because of the assumed horizontal isotropy of the reference solar model the normal-mode frequencies are independent of the direction of $\bk$. The displacement eigenfunctions may be decomposed into vertical and horizontal components according to
\begin{eqnarray}
\bxi^n \, (\br,\bk) = [ \hat{\bz} \,\xi^n_z(z,k)  + {\rm i} \,\bkhat\, \xi^n_h(z,k)]\, e^{{\rm i}\,\bk\cdot\xb},
\end{eqnarray}
where $\xb$ is a horizontal position vector and $z$ denotes height. Because the only restoring forces are pressure and buoyancy there is no motion in the direction $\hat{\bk}\times\hat{\bz}$. We choose to normalise the eigenfunctions according to
\begin{eqnarray}
\int\, \rho_0\, \bxi^{n*}(\br,\bk) \cdot \bxi^{n^\prime}(\br,\bk^\prime) \, \,d^3 \br = \de_{n, n^\prime} \, \de(\bk-\bk^\prime),
 \end{eqnarray}
where $\de_{n, n^\prime}$ is a Kronecker delta, $\de(\bk-\bk^\prime)$ is a Dirac delta function. The integral is three dimensional and over all space. For the horizontally homogeneous medium under study, we define the Green's vector $\bG^i(\xb-\bbs,z,t-\ts,\zs)$ as the displacement vector at position $\br=(\xb,z)$ and time t that results from a delta function source at position $\brs=({\mathbf s},\zs)$ and time $\ts$ pointing in the $\hat{\bbe}_i $ direction, where  $\hat{\bbe}_i $ is one of the unit vectors $\hat{\bbe}_x $, $\hat{\bbe}_y $ or $\hat{\bbe}_z $. 
If we take the Fourier transform in time, then we get
\begin{eqnarray}
 \rho_0 [ -(\om +i\,\gamma)^2 +\cH] \bG^i(\xb-\bbs,z,\om,\zs) = \frac 1{2 \pi} \de(\br -\brs) \, \hat{\bbe}_i \label{grenneqxxx}.
\end{eqnarray}

The  function $\gamma$ damps the waves so their lifetimes are finite. Typically it is chosen such that the subsequent line-widths of the model reproduce the line-widths observed in the Sun.

We now expand the $\bG^i$ on a basis of undamped normal-mode solutions according to
\begin{eqnarray}
\bG^i(\xb-\bbs,z,\om, \zs) = \sum_n \int d^2 \bk \,c_n^i(\om, \brs, \bk) \, \bxi^{n}(\br,\bk).
\end{eqnarray}
where $ c_n^i(\om, \br_s, \bk)$ are coefficients to be determined. The sum is over all radial orders $n$ and the integral is two dimensional over all horizontal wave numbers.  After taking the dot product of Equation (\ref{grenneqxxx}) with $\bxi^{n^\prime *}(\br,\bk^\prime)$ and integrating over $\br$, we find
\begin{eqnarray}
  [ -(\om +{\rm i}\,\gamma_{n^\prime} (k))^2 +\om_{n^\prime}^2(k)] c_{n^\prime}^i(\om, \brs, \bk) = \frac1{2 \pi} \xi^{n^\prime*}_i(\brs,\bk),
\end{eqnarray}
where $\gamma(k) \, \bxi^n(\br,\bk)  = \gamma_n (k) \bxi^n(\br,\bk)$. Finally, after some additional manipulation we have
\begin{equation}
  \bG^i(\bk,z,\om, \zs) = \frac 1{2\pi} \sum_n \frac{ \bzeta^n(z,\bk)\, \zeta^{n*}_i(\zs,\bk) }{ -[\om +{\rm i}\,\gamma_n(k)]^2 +\om_n^2(k)},
\label{eq.final_G}
\end{equation}
where we introduce the convenient notation for the eigenfunction
\begin{equation}
\bzeta^n(z,\bk) := \hat\bz  \, \xi^n_z(z,k)  +{\rm i} \,\bkhat\, \xi^n_h(z,k).
\label{eq.define_zeta_for_green}
\end{equation}

Equations (\ref{eq.final_G}) and (\ref{eq.define_zeta_for_green}) give a convenient set of expressions for computing the Green's functions for the damped problem in terms of the modes of the undamped problem. They are, in practice, easier to use than the expressions of \citet{2004ApJ...608..580B}, which involve the modes of the damped problem; the modes of the undamped problem are easier to compute than the modes of the damped problem. In addition, the damping model (i.e., the choice of the $\gamma_n$) can be updated independently of the computation of the normal modes.

\begin{figure}[htbp]
\begin{center}
\includegraphics[width=4.7in]{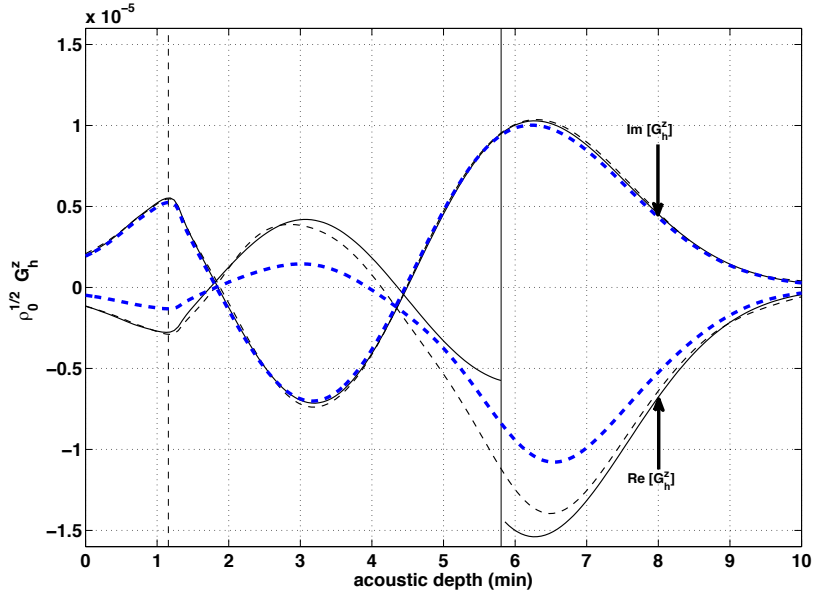}
\caption{Horizontal component of the Green's function $\bG^z$ from this study (blue dashed curves) compared with normal-mode summation Green's function from \cite{2004ApJ...608..580B} (dashed black curves) and the numerical solution  from \cite{2004ApJ...608..580B} (solid black curves). There are two curves for each Green's function to show the real and imaginary parts. The source depth is $z_s = 3.7$~Mm and the horizontal wavenumber is $k=1$~rad\,Mm$^{-1}$. The mode summation formulae uses modes from $n=0$ up to $n=14$. The vertical black dashed line shows the observation height and the vertical solid black line shows the source depth.
\label{fig.cut_greens}}
\end{center}
\end{figure}

Figure \ref{fig.cut_greens} shows how the complex Green's function presented here compares with two computations from \cite{2004ApJ...608..580B} versus acoustic depth, at fixed $\omega$ and $k$. \cite{2004ApJ...608..580B} used a slightly different definition of the Green's function: $(\partial_t + \Gamma)^2$ is replaced by $\partial_t^2 + 2\Gamma\partial_t$ in the wave operator $\mathcal{L}_0$ and $\Gamma=\Gamma(z)$. The problem was solved by \cite{2004ApJ...608..580B} by summation over the modes of the damped problem (unlike in the present paper) and also by direct numerical solution of the differential equations. The source depth is $z_s = 3.7$~Mm and the horizontal wavenumber is $k=1$~rad\,Mm$^{-1}$. For the 2004 paper, the cyclic frequency is $\omega/2\pi =3.92$~mHz, a frequency just above the $n=2$ resonance. The normal mode eigenfrequencies from the model used in this paper are slightly different from those in \cite{2004ApJ...608..580B}; on the order of a few $\mu$Hz. Thus in order to make a good comparison the new Green's function was evaluated a few $\mu$Hz higher at $\omega/2\pi=3.92265$ mHz, in order to be a similar distance from the $n=2$ mode. We find that the two mode-summation solutions (both used radial orders $0\le n\le 14$) give similar answers (dashed curves in Figure~\ref{fig.cut_greens}): The imaginary parts of the Green's functions match very well, the real parts less so. The difference in the real parts is due to the non-vanishing imaginary component of the damped eigenfunctions from \citet{2004ApJ...608..580B}. The real part of the numerical Green's function (solid curves in Figure~\ref{fig.cut_greens}) has a discontinuity at the source depth that cannot be reproduced by a sum over only 15 modes.  Whether the above differences are important for the interpretation of helioseismic travel times deserves to be studied.

\subsection{Background Solar Model and Zero Order Wavefield}

We use model S \citep{1996Sci...272.1286C} for the background model of the Sun, where sound-speed, density, pressure, and gravity are functions of height only. There are no flows or magnetic fields in this background model.

The reference displacement field for the undamped problem (i.e. $\bxi_0$) is computed by solving Equation (\ref{Lxi_eq_S}). We follow the work of \cite{2002ApJ...571..966G} and \cite{2004ApJ...608..580B} and use a plane-parallel (Cartesian) approximation. The Cowling approximation is also enforced. This reduces to an eigenvalue problem similar to that encountered in \cite{2004ApJ...608..580B}, however here we solve for $\bxi_0$ rather than $\partial_t^2 \bxi_0$. It was particularly important to ensure the eigenfunction solutions were well behaved under double numerical differentiation over height, since some of the features of the pressure kernels are sensitive to small errors. We use the same source covariance as in \cite{2004ApJ...608..580B}.

The undamped reference displacement field is used to compute the attenuated Green's function through Equation (\ref{eq.final_G}).

\subsection{Numerical resolution in Fourier space and spatial computational domain}

The kernels were computed using the Model-S non-uniform height grid that is provided as part of the Aarhus adiabatic pulsation package. We selected to use a subset of the height domain from $z=0.4959$~Mm to $z=-20.9066$~Mm. We use a uniform horizontal grid ($\Delta x = \Delta y = 96$~km) spanning 96 Mm in each direction. The size of the computational box was selected to ensure the kernels essentially vanish on the horizontal boundaries and also the bottom boundary. In particular, the amplitudes of the hyperbolic features of the kernels are notorious for decaying slowly with distance. We conducted resolution tests to confirm the example kernels presented in this paper did not change significantly with higher resolution.

Following \citet{2004ApJ...608..580B}, we used the line-width measurements from \cite{2004ApJ...602..481K} as a basis for the empirical damping model. These line widths set the numerical resolution in Fourier space that is required in the calculation of the kernels. For the $p_1$-mode examples shown in this paper, we used frequency and wavenumber resolutions of $\Delta \nu=3.8~\mu$Hz and $\Delta k = 1.59\times10^{-4}$~rad~Mm$^{-1}$ respectively. The frequency domain ranged from $f=1.5$~mHz to $f=5.3$~mHz. The wavenumber domain was from $k=0.11$~rad~Mm$^{-1}$ to $k=2.5$~rad~Mm$^{-1}$. We computed higher resolution tests to confirm that the resolution used here was sufficient.

To obtain a good representation of $p_1$ modes (including the line asymmetry), normal modes from $n=0$ to $n=14$ were used in the summation for the Green's functions. A filter was then applied to isolate the $p_1$ ridge. We used a Modulation Transfer Function (MTF) of the form $e^{-\beta k}$ where $\beta=1.15$ Mm.

\subsection{Example kernels}

The code used to compute travel-time sensitivity kernels for this paper was written in MATLAB and is a modified and extended version of the code used by \cite{2004ApJ...608..580B} to compute mean sound-speed kernels and by \cite{2007AN....328..228B} to compute difference velocity kernels. The main modifications are the computation of density and pressure kernels and the new method to compute Green's functions (\S\ref{sec.greens}). For simplicity, when computing the density kernel we neglected the perturbation the gravitational field ($\delta \bg$) and $\gamma$ in Equation (\ref{eq.def_curlyc_rho}); both these effects are small.

\begin{figure}[htbp]
\begin{center}
\includegraphics[width=3.8in]{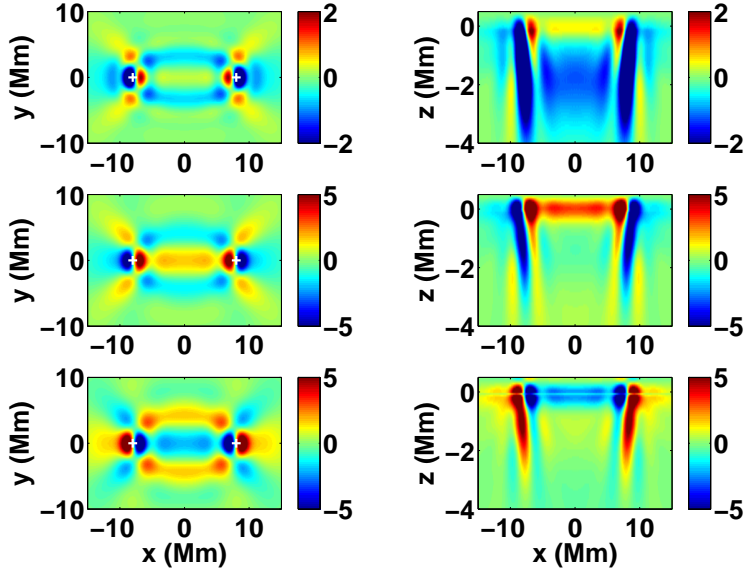}
\caption{Mean point-to-point travel-time sensitivity kernels for relative changes in sound-speed squared (top panels), density (middle panels), and pressure (bottom panels). Kernels have units of s Mm$^{-3}$. Normal modes $n=0$ to $n=14$ were used to compute the Green's function and a filter was applied to isolate the $p_1$-mode ridge. The left column displays horizontal cuts at a height of $z=-0.3$ Mm. The right column displays vertical slices along the line $y=0$ Mm. All colour scales are symmetric about zero (green) with red positive and blue negative. The crosses in the left panels represent the two points $\xb_1$ and $\xb_2$; the distance between the two points is \thedelta~Mm.\label{fig.Ksca_mean_p1}}
\end{center}
\end{figure}

Figure~\ref{fig.Ksca_mean_p1} shows horizontal and vertical slices through kernels for the changes in the mean travel-time due to fractional perturbations to the squared sound-speed, density, and pressure (for $p_1$-modes) and a separation of $\Delta=\xb_2-\xb_1=\thedelta \; {\rm Mm}\;\hat{\xb}$ between the observation points. All the kernels are symmetric in both horizontal coordinates and show the same general features as the sound-speed kernels shown by \citet{2004ApJ...608..580B}. The kernels have complicated features and are largest in amplitude near the two observation points. The kernels extend roughly over the first few Mm below the photosphere; this is the region where the $p_1$ kinetic energy density is significant for frequencies 1.5-5.3 mHz. The total integral of the sound-speed kernel is negative (increased sound-speed leads to a reduction in $\dtmn$). The total integral of the density kernel is also negative implying a uniform increase in density would also lead to a reduction in $\dtmn$. Conversely, the total integral of the pressure kernel is positive, which implies a uniform increase in pressure would lead to an increase in $\dtmn$. The pressure and density kernels are strongly anti-correlated.

\begin{figure}[htbp]
\begin{center}
\includegraphics[width=3.8in]{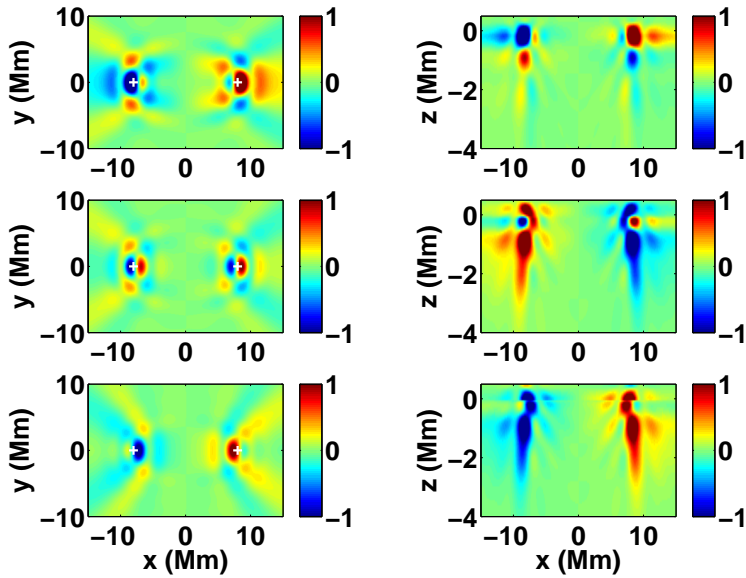}
\caption{Difference point-to-point travel-time sensitivity kernels for relative changes in sound-speed squared (top panels), density (middle panels), and pressure (bottom panels). Kernels have units of s Mm$^{-3}$. Normal modes $n=0$ to $n=14$ were used to compute the Green's function and a filter was applied to isolate the $p_1$-mode ridge. The left column displays horizontal cuts at a height of $z=-0.3$ Mm. The right column displays vertical slices along the line $y=0$ Mm. The crosses in the left panels represent the two points $\xb_1$ and $\xb_2$; the distance between the two points is \thedelta~Mm.\label{fig.Ksca_diff_p1}}
\end{center}
\end{figure}

Figure~\ref{fig.Ksca_diff_p1} shows slices through the kernels for the sensitivity of the travel-time differences to changes in sound speed, density, and pressure, for the same geometry and filter as the kernels shown in Figure~\ref{fig.Ksca_mean_p1}. These kernels, unlike in Figure~\ref{fig.Ksca_mean_p1}, are anti-symmetric in $x$, but still symmetric in $y$. This is a consequence of the definition of the travel-time difference (the difference between the two one-way travel times). The anti-symmetry in $x$ is important as it implies that the total integral of these kernels vanishes, i.e., a spatially uniform change in solar structure does not produce a change in the travel-time differences.

\begin{figure}[htbp]
\begin{center}
\includegraphics[width=3.8in]{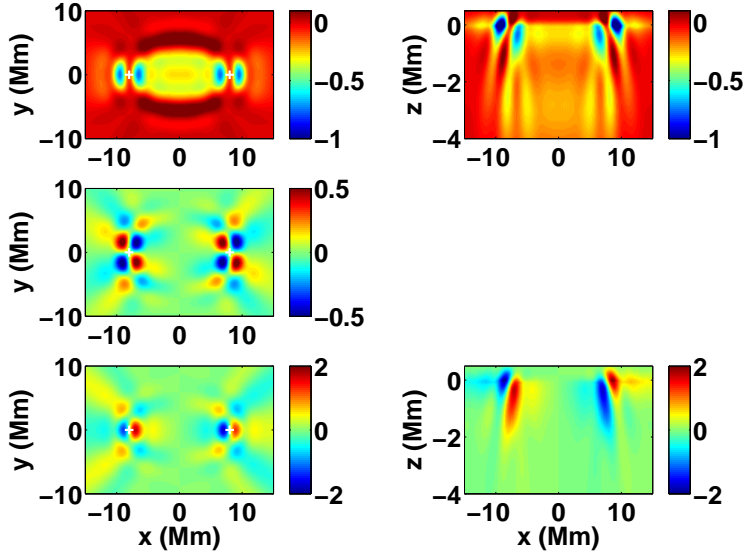}
\caption{Difference point-to-point travel-time sensitivity kernels for $v_x$ (top panels), $v_y$ (middle panels), and $vz$ (bottom panels). Kernels have units of s Mm$^{-3}$ / (km/s). Normal modes $n=0$ to $n=14$ were used to compute the Green's function and a filter was applied to isolate the $p_1$-mode ridge. The left column displays horizontal cuts at a height of $z=-0.3$ Mm. The right column displays vertical slices along the line $y=0$ Mm. All colour scales are symmetric about zero (green) with red positive and blue negative. The crosses in the left panels represent the two points $\xb_1$ and $\xb_2$; the distance between the two points is \thedelta~Mm.\label{fig.Kvel_diff_p1}}
\end{center}
\end{figure}

Figure~\ref{fig.Kvel_diff_p1} shows slices through the kernels for the sensitivity of the travel-time differences to changes in $v_x$, $v_y$, and $v_z$, for the same geometry and filter as the kernels shown in Figure~\ref{fig.Ksca_mean_p1}. The kernel for $v_z$ is anti-symmetric in $x$ and symmetric in $y$, whereas the kernel for $v_y$ is anti-symmetric in both $x$ and $y$. This implies that uniform perturbations in $v_z$ and $v_y$ will not effect travel-time differences. The difference kernel for $v_x$ is symmetric in both $x$ and $y$ and consequently has a non-zero total integral. 3D difference kernels for the f-mode was presented in \citet{2007AN....328..228B} and \citet{2011AN....332..658B}. The present calculations are fully consistent with \citet{2007AN....328..228B} and \citet{2011AN....332..658B}.

\begin{figure}[htbp]
\begin{center}
\includegraphics[width=3.8in]{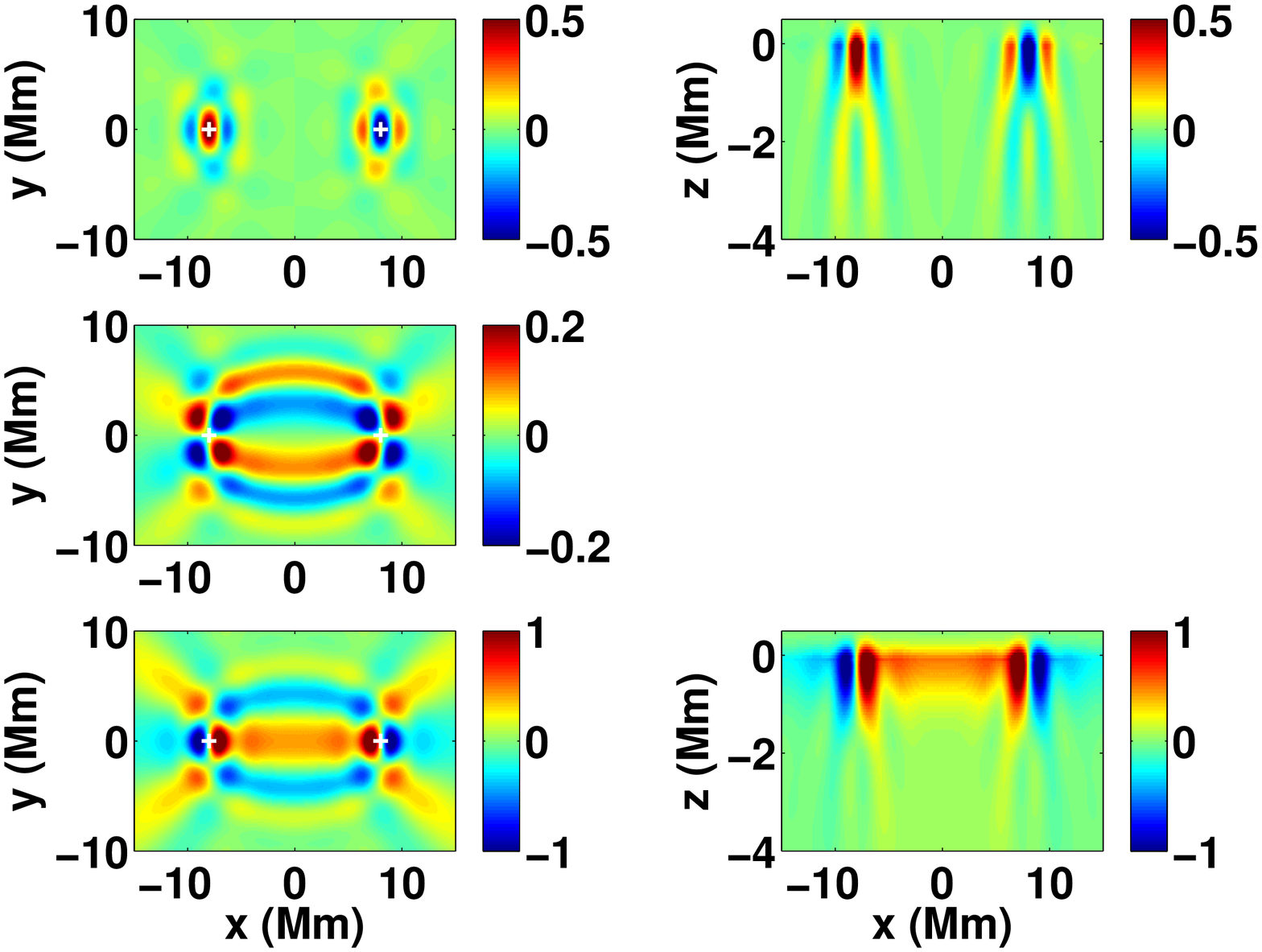}
\caption{Mean point-to-point travel-time sensitivity kernels for $v_x$ (top panels), $v_y$ (middle panels), and $vz$ (bottom panels). Kernels have units of s Mm$^{-3}$ / (km/s). Normal modes $n=0$ to $n=14$ were used to compute the Green's function and a filter was applied to isolate the $p_1$-mode ridge. The left column displays horizontal cuts at a height of $z=-0.3$ Mm. The right column displays vertical slices along the line $y=0$ Mm. All colour scales are symmetric about zero (green) with red positive and blue negative. The crosses in the left panels represent the two points $\xb_1$ and $\xb_2$; the distance between the two points is \thedelta~Mm.\label{fig.Kvel_mean_p1}}
\end{center}
\end{figure}

Figure~\ref{fig.Kvel_mean_p1} shows slices through the kernels for the sensitivity of mean travel-times to changes in $v_x$, $v_y$, and $v_z$, for the same geometry and filter as the kernels shown in Figure~\ref{fig.Ksca_mean_p1}. The symmetries imply that uniform perturbations in both $v_x$ and $v_y$ would not affect mean travel times. Equivalent 2D kernels for $v_x$ and $v_y$, for the f-mode, were shown in \citet{2007ApJ...671.1051J}. We show here that the mean kernel for $v_z$ is symmetric in both $x$ and $y$ and has a significant non-zero total integral.

\begin{figure}[htbp]
\begin{center}
\includegraphics[width=4.7in]{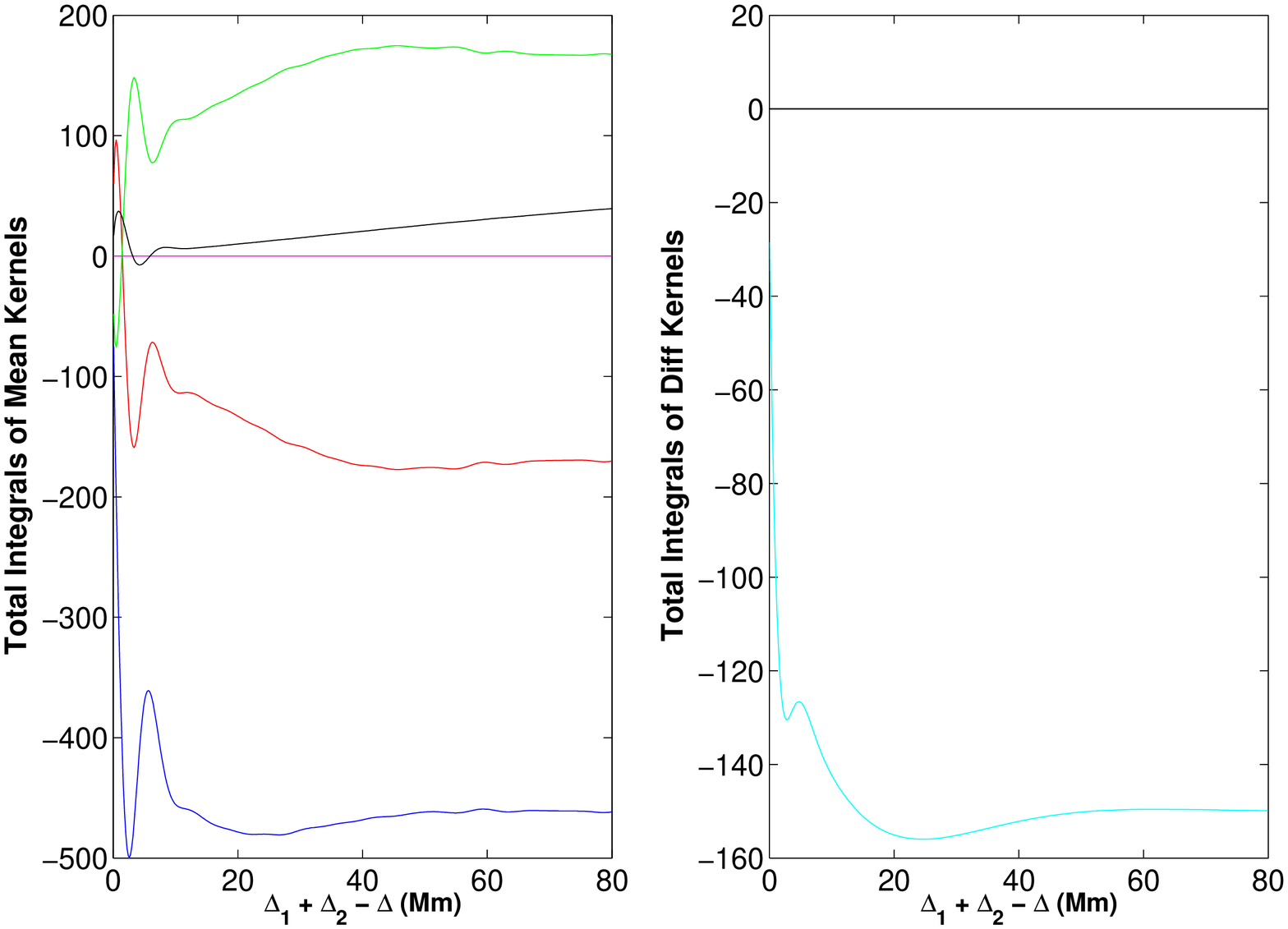}
\caption{Total 3D integrals of the kernels as a function of $\Delta_1 + \Delta_2 -\Delta$ (Mm). The flow kernels were scaled by a uniform flow of 1 km/s such that all the total integrals have units of seconds. The left panel is for mean kernels and the right panel for difference kernels. The integrals for sound-speed squared is blue, density is red, pressure is green, $v_x$ is cyan, $v_y$ is magenta, and $v_z$ is black. The total integrals for the mean kernels for sound-speed squared, density and pressure are close to converged at large distances from the two points however, for the vertical velocity a larger domain is necesssary to ensure convergence. The total integrals of the mean kernels for horizontal flows vanish. The only non-zero total integral for the difference kernels is the one for $v_x$ (cyan) and its total integral is close to converged with the domain of 96 Mm.
\label{fig.TI_vs_Lh}}
\end{center}
\end{figure}

Figure \ref{fig.TI_vs_Lh} demonstrates how the total integrals of the kernels converge as a function of distance from the observation points. The total integrals for the structure and horizontal flow kernels have approximately converged over the 96 Mm$^2$  domain, however, the mean kernel for vertical velocity has not yet converged.

Table \ref{tab.total_integrals} shows the mean and difference total integrals for $p_1$ modes. This gives an indication of how travel-time shifts respond to uniform perturbations. The largest contribution to the mean travel-times is from a uniform change in relative sound-speed. The contributions to mean travel-times from relative density and pressure perturbations are almost equal but opposite in sign. For the flows, the only non-zero contribution to the mean travel-times would be a uniform vertical flow, the horizontal components vanish due to their symmetries. The only contribution to the difference travel times due to uniform perturbations is due to flows along the two observation points.
\begin{table}
\begin{center}
\begin{tabular}{| c | c | c | c | c | c | c | }
\multicolumn{7}{c}{}\\
\hline
 & $\int K_{c^2} d \br$ & $\int K_{\rho} d\br$ & $\int K_P d \br$ & $\int K_{v_x} d \br$ & $\int K_{v_y} d \br$ & $\int K_{v_z} d \br$\\
\hline
mean & -460 s & -170 s & 168 s & 0 s/(km/s) & 0 s/(km/s) & 54 s/(km/s)  \\
\hline
diff & 0 s & 0 s & 0 s & -152 s/(km/s) & 0 s/(km/s) & 0 s/(km/s) \\
\hline
\end{tabular}
\caption{Total integrals of the mean and difference kernels for sound-speed, density, pressure and flows for $p_1$ modes.}
\label{tab.total_integrals}
\end{center}
\end{table}

\section{A consistency check}

In this section, we compute the change in the cross-covariance (e.g. see Equation (\ref{eq.xcorr_def})), $\Delta C$, caused by a simple sound-speed perturbation of the form
\begin{equation}
\frac{\Delta c^2}{c^2_0}  = \epsilon \exp{\left[\frac{-(z-z_0)^2}{2 \sigma^2} \right]},
\end{equation}
where $\epsilon = 0.0001$, $z_0 =-1.5$ Mm, and $\sigma = (1\;\mbox{Mm}) / (8 \ln 2)^{1/2}$. 

Because $\epsilon$ is very small, $ \Delta C$ can be approximated by
\begin{eqnarray}\label{xcorr2}
\delta C(\omega)   = \int d \xb \, dz \,\, \cC_{c^2}(\xb, z, \om; \xb_1, \xb_2) \frac{\delta c^2}{c_0^2}(z),
\end{eqnarray}
where the two points are separated by a distance of $16$~Mm. The function $\delta C(\omega)$ is real.

\begin{figure}[h]
\begin{center}
\includegraphics[width=4.8in]{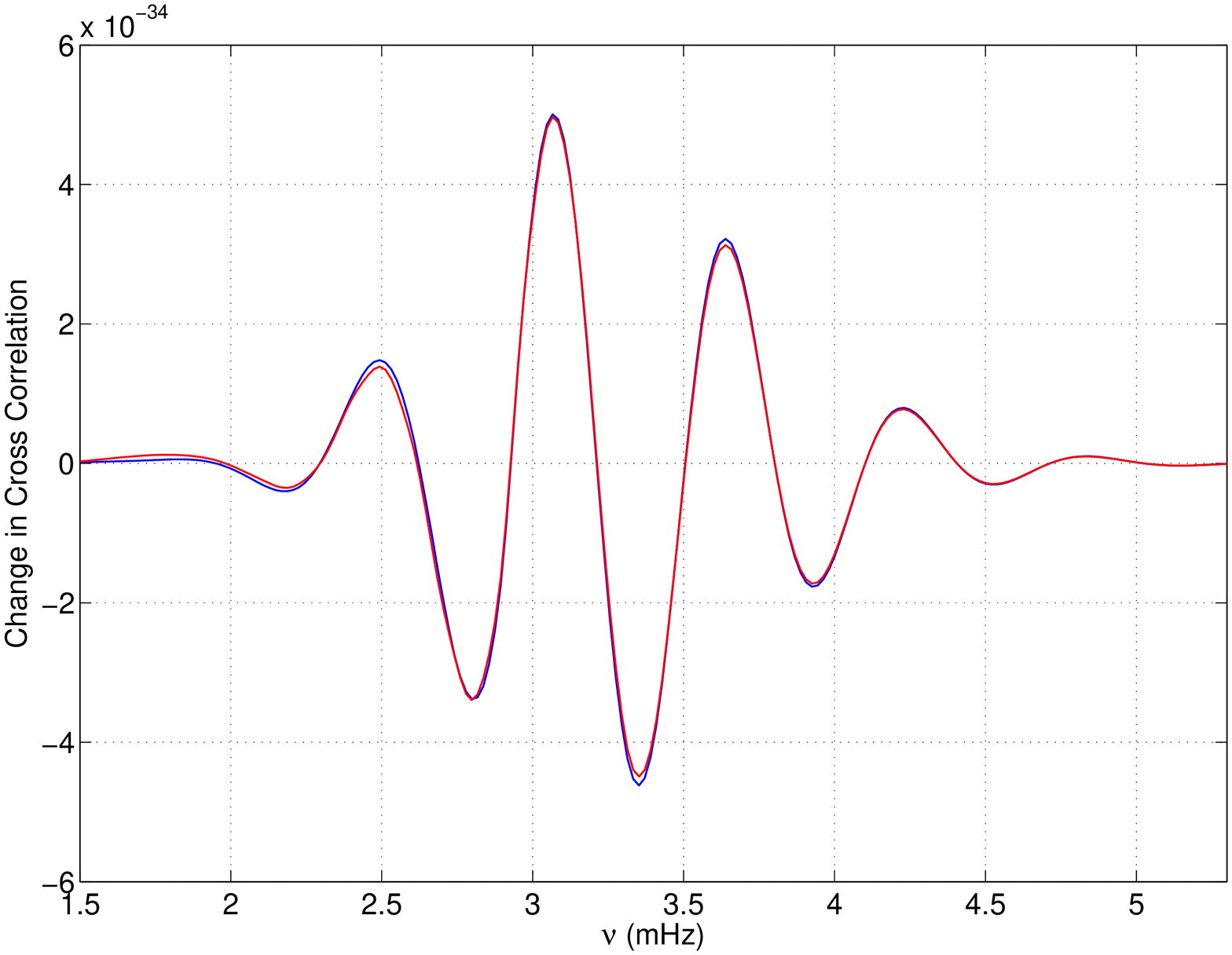}
\caption{Change in the cross-covariance (at a fixed point-to-point distance of $\Delta = 16.0347$ Mm) for the case of the Gaussian perturbation to the sound-speed of Model S. The red curve is $\delta C$ and the blue curve $\Delta C$. The results show very good agreement.}
\label{fig.delta_C_due_to_c}
\end{center} 
\end{figure}

Because $\Delta c^2$ depends on height only, we can also compute $\Delta C$ using
\begin{eqnarray}\label{xcorr1}
\Delta C(\omega)     = C(\omega)   - C_0(\omega).
\end{eqnarray}
where $C(\omega)$ is the cross-covariance computed using the eigenmodes of the solar model with sound speed $c_0+\Delta c$. To compute the new eigenmodes we use the same solver as in Section~2.2. Here we used a different damping model where each mode was damped according to $\gamma_n(k) = \tilde\gamma \,|\omega_0(k)/\tilde\omega|^{4.4}$, where $\tilde\omega/(2\pi) = 3 $ mHz and $\tilde\gamma/(2\pi)=100$ $\mu$Hz. In addition, for wave numbers less than 0.8 Mm we set the damping to remain constant. The resolution in wavenumber was the same as earlier, however the resolution in frequency was changed to $19.1$ mHz. In Figure~\ref{fig.delta_C_due_to_c}, we find that there is a very good agreement between $\Delta C(\omega)$ and $\delta C(\omega)$, which is expected in the limit $\epsilon \rightarrow 0$.  These results  give us confidence in the reliability of the computations. Furthermore, it appears that the numerical resolutions and the extent of the horizontal domain are sufficient.

\section{Are there simple relationships between the kernels?}

\begin{figure}[htbp]
\begin{center}
\includegraphics[width=4.8in]{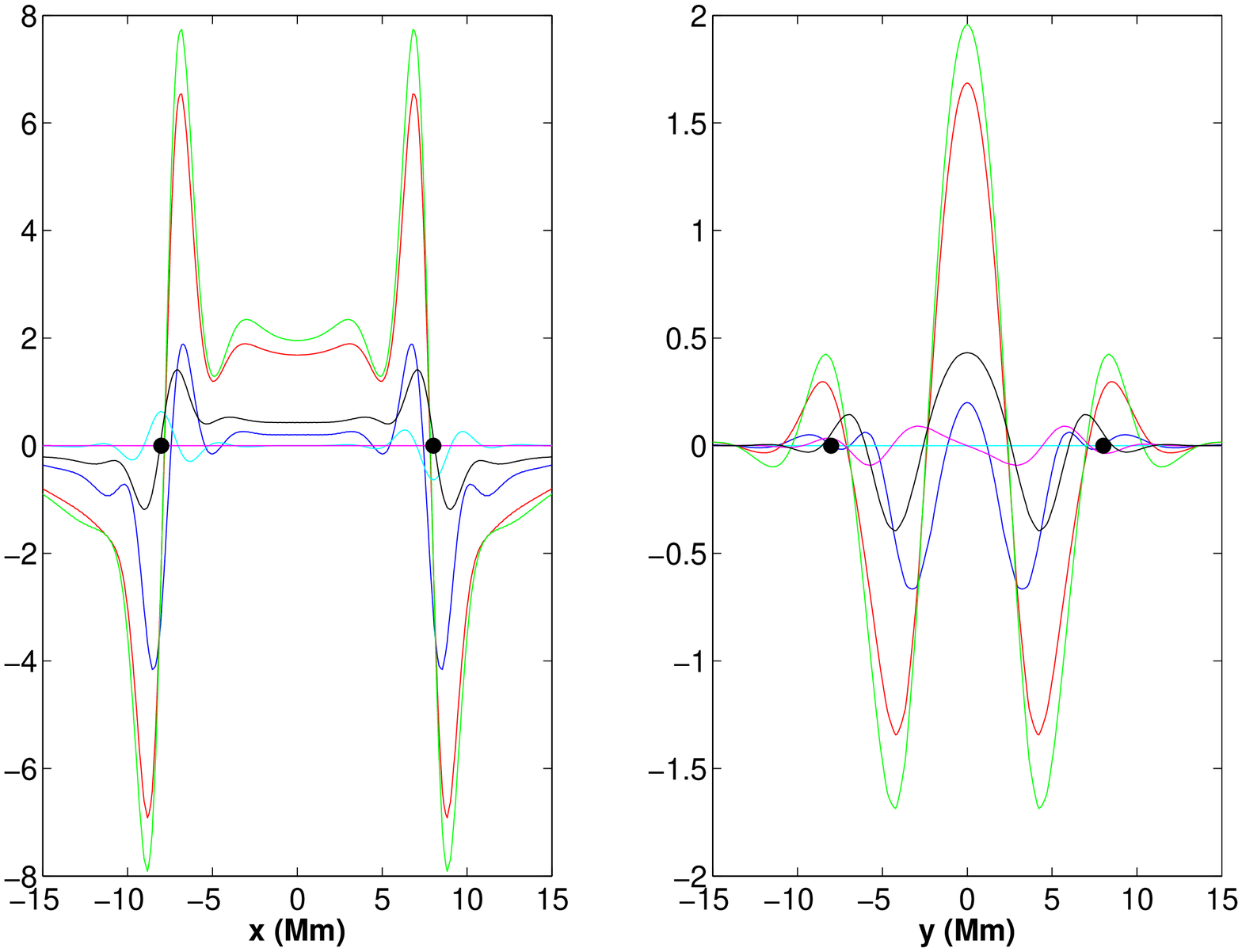}
\caption{Cuts through the mean point-to-point travel-time sensitivity kernels for sound-speed squared (blue), density (red), negative pressure (green), $v_x$ (cyan), $v_y$ (magenta), and $vz$ (black). The sound-speed, density and pressure kernels have units of s Mm$^{-3}$, whereas the velocity kernels have units Mm$^{-3}$ / (km/s). Normal modes $n=0$ to $n=14$ were used to compute the Green's function and a filter was applied to isolate the $p_1$-mode ridge. The cuts are at taken at a height of $z=-0.3$ Mm. The left panel is a cut along the x axis, whereas the right panel is a cut along the y axis. The distance between the two points is \thedelta~Mm.\label{fig.kernel_mean_cuty0x0}}
\end{center}
\end{figure}

\begin{figure}[htbp]
\begin{center}
\includegraphics[width=4.8in]{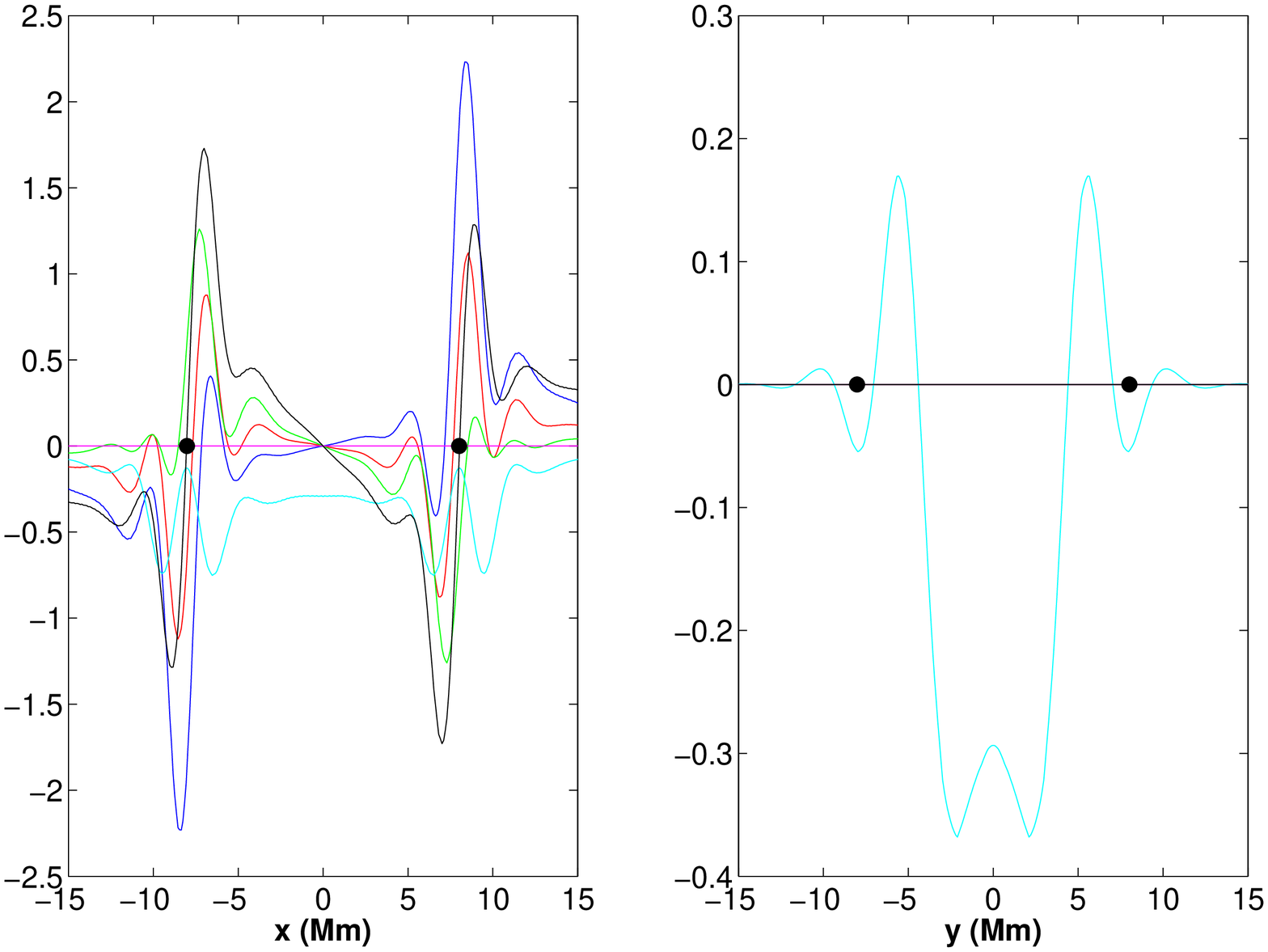}
\caption{Cuts through the difference point-to-point travel-time sensitivity kernels for sound-speed squared (blue), density (red), negative pressure (green), $v_x$ (cyan), $v_y$ (magenta), and $vz$ (black). The sound-speed, density and pressure kernels have units of s Mm$^{-3}$, whereas the velocity kernels have units Mm$^{-3}$ / (km/s). Normal modes $n=0$ to $n=14$ were used to compute the Green's function and a filter was applied to isolate the $p_1$-mode ridge. The cuts are at taken at a height of $z=-0.3$ Mm. The left panel is a cut along the x axis, whereas the right panel is a cut along the y axis. The distance between the two points is \thedelta~Mm.\label{fig.kernel_diff_cuty0x0}}
\end{center}
\end{figure}

In order to produce reliable time-distance inversions of the solar interior it is helpful if the kernels have different structures and features. Figures \ref{fig.kernel_mean_cuty0x0} and \ref{fig.kernel_diff_cuty0x0} display cuts through the mean and difference kernels at $z=-0.3$ Mm along the x and y axes. We have plotted the negative values of the pressure kernel (green) in order to better illustrate the strong anti-correlation with the kernel for density (red). It can evident that kernels for sound-speed squared (blue), density (red) and negative pressure (green) have similar features.

Travel-time shifts due to sound-speed perturbations are directly related to the divergence of the eigenfunctions. Therefore, travel-times shifts for f-modes are not sensitive to sound-speed perturbations since the divergences of the f-mode eigenfunctions vanish. However, the power spectrum in the neighbourhood of the f-mode ridge has a small contribution from the wings of the p modes and thus f-mode mean travel-time measurements are weakly sensitive to sound-speed perturbations. For example, we find that the total integral of $K_{c^2}$ for the f-mode ridge at $\Delta=\thedelta$~Mm is approximately 3\% of the total integral of $K_{c^2}$ for the $p_1$ ridge.

\section{Specific case of hydrostratic equilibrium}\label{sec.hydro}

In the fully general non-hydrodynamic case, where we have three independent thermodynamic perturbation variables, the travel-time shifts (ignoring flows) are given by
\begin{eqnarray}
\delta\tau= \int_\odot  d\br \left[ K_{c^2,\rho \, P}(\br) \, \frac{\delta c^2}{c^2_0}(\br)  +  K_{\rho,c^2 P}(\br) \, \frac{\delta \rho}{\rho_0}(\br)  +  K_{P,c^2 \rho}(\br) \, \frac{\delta P}{P_0}(\br)\right].
\label{eq.dt_from_dcsq_drho_hydro}
\end{eqnarray}
We have now explicitly expressed which quantities are held fixed. The kernel $K_{c^2,\rho \, P}$ is due to perturbations in sound-speed at fixed density and pressure, similarly $K_{\rho,c^2 \, P}$ is the density kernel at constant sound-speed and pressure and finally $K_{P,c^2 \rho}$ is the pressure kernel at constant sound-speed and density.

In global helioseismology, it is typically assumed that the density and pressure perturbations are in hydrostatic equilibrium \citep[e.g][]{1991sia..book..519G}
\begin{equation}
\nab \delta P = \bg_0 \,\delta \rho \label{eq.hydro1},
\end{equation}
where we assume the perturbation to the gravitational field is negligible for the sake of simplicity.

Equation (\ref{eq.hydro1}) may be manipulated to express the pressure perturbation entirely in terms of a density perturbation (or vice-versa) and thus the number of independent thermodynamic variables is reduced from three to two.

The horizontal components of Equation (\ref{eq.hydro1}) implies that that the pressure perturbation is at most a function of height, i.e. $\delta P = \delta P(z)$. The vertical component implies that the density perturbation is also at most a function of height i.e $\delta \rho = \delta \rho(z)$, and hence
\begin{eqnarray}
\frac{d \delta P}{dz} &=& -g_0 \,\delta \rho \label{eq.dpz}.
\end{eqnarray}

It is possible to implement the assumption of hydrostatic equilibrium, for example by eliminating pressure in terms of density. To achieve this it is necessary to manipulate the last term on the right-hand-side of  Equation (\ref{eq.dt_from_dcsq_drho_hydro}) as a gradient of the pressure. So since the perturbations are all functions of height only, we require
\begin{equation}
\frac{\partial A(\br)}{\partial z} = \frac{K_{P,c^2 \rho}(\br) }{P_0(z)} 
\end{equation}
with boundary condition $A(\xb,z_t) =0$. This could be solved for each $\xb$. Provided a solution exists, after integration by parts once, then the travel time shifts become
\begin{eqnarray}
\delta\tau= \int_\odot  d\br \left[ K_{c^2,\rho \, P}(\br) \, \frac{\delta c^2}{c^2_0}(z)  +  \left(K_{\rho,c^2 P}(\br) +\rho_0(z) g_0(z) A(\br) \right) \, \frac{\delta \rho}{\rho_0}(\br)  \right],
\end{eqnarray}
However, since we know we have two independent thermodynamic quantities then it is also possible to express the travel time shifts in terms of a different set of kernels according to
\begin{eqnarray}
\delta\tau= \int_\odot  d\br \left[ K^H_{c^2,\rho}(\br) \, \frac{\delta c^2}{c^2_0}(z)  +  K^H_{\rho,c^2}(\br) \frac{\delta \rho}{\rho_0}(z)  \right],
\end{eqnarray}
where the superscript $H$ is to remind the reader that the assumption of hydrodynamic balance has been asserted. Then by equating coefficients we get
\begin{eqnarray}
K^H_{c^2,\rho}(\br)  &=& K_{c^2,\rho P}(\br)   \label{eq.k_hydro_nonhydro1}\\
K^H_{\rho,c^2}(\br)  &=& K_{\rho,c^2 P}(\br) +\rho_0(z) g_0(z) A(\br) \label{eq.k_hydro_nonhydro2}
\end{eqnarray}
Equation (\ref{eq.k_hydro_nonhydro1}) implies that the sound-speed kernel at constant density and pressure in the general problem without hydrostatic equilibrium is equal to the sound-speed kernel at constant density in the hydrostatic case. Equation (\ref{eq.k_hydro_nonhydro2}) demonstrates that more work is required to obtain the density kernel at constant sound-speed in the hydrostatic case. It is related to the density kernel at constant sound-speed and pressure in the non-hydrodynamic case plus an additional term involving the function $A(\br)$. Most of the work in going from one set of kernels to another involves computing $A(\br)$. 

\section{Discussion}

We have reviewed the sensitivity of time-distance travel times to small-amplitude steady changes in the structure of a solar model as well as their sensitivity to weak and steady flows. In the course of this paper, we introduced a new convenient method for computing Green's functions based on normal-mode summation using the normal modes of the adiabatic wave equation \citep[in contrast with the approach of][based on the modes of the damped problem]{2004ApJ...608..580B}. The sensitivity kernels we computed are essential for time-distance inversions of the solar interior.

It is difficult to disentangle density and pressure perturbations and this will complicate time-distance inversions. Our calculations have shown that, to a good approximation, there exists a constant $\alpha > 0$, such that  $K_{\rho, cP} \simeq - \alpha K_{P, c\rho}$. We have
\begin{equation}
 \delta \tau = \int dV \left( K_{c^2,P\rho} \frac{\delta c^2}{c^2} + K_{P,c\rho} \frac{\delta P}{P}  + K_{\rho,cP} \frac{\delta \rho}\rho \right)    \simeq      \int dV \left( K_{c^2,u} \frac{\delta c^2}{c^2} +  K_{u,c} \frac{\delta u}{u} \right) 
\end{equation}
where $u \equiv P/\rho^\alpha$, $K_{c^2,u} \equiv K_{c^2, P \rho}$, and $K_{u,c} \equiv K_{P,c\rho}$. Thus, to a good approximation, two thermodynamic quantities are enough to describe the medium. If the density and pressure perturbations are in hydrostatic equilibrium, then two thermodynamic quantities are exactly enough (see Section \ref{sec.hydro}).

The correct treatment of instrumental, geometrical, and mode physics effects is an important future goal for accurately imaging the solar interior. Correctly including line-of-sight and foreshortening effects in the computation of the kernels, will allow accurate applications using large line-of-sight angles. \citet{2007ApJ...671.1051J} studied line-of-sight effects for 2D kernels (for surface waves) and found striking changes to the appearance of kernels that will be important for accurately inverting for small-scale flows. Computing travel-time sensitivity kernels in full spherical geometry, rather than using the Cartesian approximation, is important for larger distances between the observation points.

The kernels discussed in this review assume that the Lagrangian pressure perturbation vanishes at the top boundary. This is not suitable for frequencies above the acoustic cut-off frequency and thus needs to be improved. Another important question is how much do time-distance inversion results depend on accurate descriptions of input physics, such as source models or damping.

Most cases assume a laterally homogeneous reference (non-magnetic) model. To study more complex features on the Sun (e.g. sunspots), kernels with laterally heterogeneous reference models are important. Initial steps towards this goal has already been achieved by \cite{2011ApJ...738..100H,2012PhRvL.109j1101H}.

\begin{acknowledgements}
This work was carried out in the framework of Collaborative Research Center SFB 963 ''Astrophysical Flow Instabilities and Turbulence" (Project A1) from the German Science Foundation (DFG). L.G. acknowledges support from EU FP7 Collaborative Project ''Exploitation of Space Data for Innovative Helio- and Asteroseismology'' (SPACEINN) and ERC Starting Grant ''Seismic Imaging of the Solar Interior''.
\end{acknowledgements}

\bibliographystyle{aps-nameyear}

\end{document}